\documentclass[12pt]{amsart}
\usepackage{amssymb, amsmath}
\usepackage{fourier}

\theoremstyle{plain}

\newtheorem{theorem}{Theorem}

\newtheorem{step}{Step}

\newtheorem{definition}[theorem]{Definition}

\newtheorem{proposition}[theorem]{Proposition}

\theoremstyle{remark}
\newtheorem*{Remark}{Remark}
\newtheorem{example}{Example}

\newcommand{\bit}{\begin{itemize}}
\newcommand{\eit}{\end{itemize}}
\newcommand{\ben}{\begin{enumerate}}
\newcommand{\een}{\end{enumerate}}
\newcommand{\be}{\begin{equation}}
\newcommand{\ee}{\end{equation}}
\newcommand{\ba}{\begin{array}}
\newcommand{\ea}{\end{array}}

\newcommand{\mc}[1]{\mathcal{#1}}
\newcommand{\supp}[1]{\mathrm{supp}\left(#1\right)}

\newcommand{\norm}[1]{\left|\left|#1\right|\right|}

\newcommand{\eps}{\varepsilon}

\newcommand{\inner}[2]{\left< #1,#2 \right>}

\newcommand\cB{\mathcal B}
\newcommand\cH{\mathcal H}
\newcommand\cP{\mathcal P}
\newcommand\cS{\mathcal S}

\newcommand\R{\mathbb R}

\newcommand{\fel}{\frac{1}{2}}

\newcommand\tr{\operatorname{Tr}}
\newcommand{\ler}[1]{\left( #1 \right)}

\begin{document}

\title[]{Quantum $f$-divergence preserving maps on positive semidefinite operators acting on finite dimensional Hilbert spaces}

\author{D\'aniel Virosztek}
\address{Department of Analysis, Institute of Mathematics\\
Budapest University of Technology and Economics\\
H-1521 Budapest, Hungary
and
MTA-DE ``Lend\" ulet'' Functional Analysis Research Group, Institute of Mathematics\\
         University of Debrecen\\
         H-4002 Debrecen, P.O. Box 400, Hungary}
\email{virosz@math.bme.hu}
\urladdr{http://www.math.bme.hu/\~{}virosz}

\thanks{
The author was supported by the ``Lend\" ulet'' Program (LP2012-46/2012) of the Hungarian Academy of Sciences and by the Hungarian Scientific Research Fund (OTKA) Reg. No.  K104206. The author was also supported by the ``For the Young Talents of the Nation'' scholarship program (NTP-EF\"O-P-15-0481) of the Hungarian State.}



\begin{abstract}
We determine the structure of all bijections on the cone of positive semidefinite operators which preserve the quantum $f$-divergence for an arbitrary strictly convex function $f$ defined on the positive halfline. It turns out that any such transformation is implemented by either a unitary or an antiunitary operator.
\end{abstract}
\maketitle
\section{Introduction}
In classical information theory and statistics the so-called \emph{$f$-divergen\-ces} --- which were introduced by \emph{Imre Csisz\'ar,} see \cite{csisz} --- are widely used to measure dissimilarity between probability distributions. For example the famous \emph{Kullback-Leibler divergence} \cite{kl} is a particular $f$-divergence. The quantum information theoretical counterpart of the notion of $f$-divergence was developed by \emph{D\'enes Petz}. In 1985, he introduced the concept of \emph{quasi-entropies} for states of a von Neumann algebra \cite{petz}. 
The quasi-entropies form a wide family of generalized distance measures. The reader who is interested in this large class of divergences should consult also \cite{Petz-orig}. The \emph{quantum $f$-divergences} are recovered from quasi-entropies with a particular choice of a particular parameter --- see \cite{HMPB} for more details.
\par

A few years ago, \emph{Moln\'ar, Nagy and Szokol} determined the structure of those (not necessarily bijective) transformations which act on the \emph{state space} of a finite quantum system and leave the quantum $f$-divergence invariant \cite{mnsz13}. The state space of an $n$-level quantum system consists of positive semidefinite operators acting on an $n$-dimensional Hilbert space with trace $1.$
Quantum $f$-divergences were originally defined on positive semidefinite operators (see, e.g., \cite{HMPB} and \cite{Petz-orig}), hence the following question naturally appears. What happens if the cone of positive semidefinite operators is taken into consideration instead of the state space. Our aim is to answer this question.
\par
In this paper we describe the structure of the preserver transformations of the set of positive semidefinite operators with respect to an arbitrary quantum $f$-divergence. The only restriction is that the generating function $f$ must be strictly convex.
\par
Let us mention some of the previous results wich are related to our work.
In the paper \cite{molord} Moln\'ar shows that if $\phi$ is a bijective map on the set of nonsingular states which preserves \emph{Umegaki's relative entropy,} then $\phi(.)=U.U^*$ for some unitary or antiunitary operator $U,$ see \cite[Thm. 3]{molord}.
Later on, Moln\'ar and Nagy provided the same result for other important sorts of divergences, namely the \emph{Tsallis relative entropy} and the \emph{quadratic relative entropy,} see \cite[Thm. 3]{molnagy}. It is proved in the same work that concerning any of the aforementioned two relative entropies or the \emph{Belavkin-Staszewski relative entropy} or the \emph{Jensen-Shannon divergence}, the preserver transformations on the \emph{whole} state space are exactly the unitary or antiunitary conjugations \cite[Thm. 2]{molnagy}.
The notion of quantum $f$-divergence may be considered as a certain generalization of the Umegaki relative entropy. Another possible generalization of Umegaki's relative entropy is the notion of \emph{Bregman $f$-divergence.} The preservers of Bregman $f$-divergences on the set of positive definite matrices and on the state space are determined in \cite{mpv15} and in \cite{dv16a}, respectively. For some other recent results on Bregman divergences the reader should consult, e. g., the papers \cite{ls14} and \cite{pv15}. 
\par
Throughout this paper the following notation will be used. $\mc{H}$ stands for a finite dimensional complex Hilbert space and $\mc{B}(\mc{H})$  denotes the set of linear operators on the Hilbert space $\mc{H}.$ Usually, $\cB(\cH)$ denotes the set of bounded linear operators on a Hilbert space $\cH.$ However, in the finite dimensional case --- which is the object of this paper --- every linear operator is bounded. The symbols $\mc{B}^{sa}(\mc{H})$ and $\mc{B}^{+}(\mc{H})$ stand for the self-adjoint and positive semidefinite operators on $\cH,$ respectively. The linear space $\mc{B}(\mc{H})$ is endowed with the Hilbert-Schmidt inner product $\inner{X}{Y}_{HS}=\tr X Y^{*}$ and $\norm{.}_{HS}$ denotes the induced norm. $\cS\ler{\cH}$ stands for the state space of $\cH$ (the set of positive semidefinite operators with unit trace) and $\cP_1(\cH)$ denotes the set of rank-one projections on $\cH.$

If $f: \, I \rightarrow \R$ is a function defined on an interval $I \subset \R$ then the corresponding \emph{standard operator function} is the following map:
$$
f: \{A \in \mc{B}^{sa}(\mc{H}): \ \sigma(A) \subseteq I \} \rightarrow \mc{B}(\mc{H})
$$
$$
A=\sum_{a \in \sigma(A)} a P_a \mapsto f(A):=\sum_{a \in \sigma(A)} f(a) P_a,
$$
where $\sigma(A)$ is the spectrum and $P_a$ is the spectral projection corresponding to the eigenvalue $a.$

\par
The following statement is folklore and can be verified by easy computations. However, we would like to refer to it in the following several times, hence we present this statement as a proposition.
\begin{proposition} \label{convmon}
Let $f: I \rightarrow \R$ be a function defined on some interval $I \subset \R.$ Then $f$ is strictly convex --- that is, $f(\lambda a+(1-\lambda)b)<\lambda f(a)+(1-\lambda)f(b)$ for any $a \neq b \in I, \, \lambda \in (0,1)$--- if and only if the corresponding difference quotient function $h(a,b)=\frac{f(a)-f(b)}{a-b}$ is strictly monotone increasing in its both variables.
 
\end{proposition}

\subsection{Quantum $f$-divergences on positive semidefinite operators}
Let $f: [0,\infty) \rightarrow \R$ be a strictly convex function. By the convexity, $f$ is necessarily continuous on $(0,\infty).$ 
The difference quotient $\frac{f(x)-f(0)}{x-0}$ is monotonically increasing, hence the limit $\lim_{x\to +\infty}\frac{f(x)-f(0)}{x-0}$ exists and is finite or $+\infty.$ Therefore, the limit $\lim_{x\to +\infty}\frac{f(x)}{x}$ exists as well, since
$$
\lim_{x\to +\infty}\frac{f(x)-f(0)}{x-0}=\lim_{x\to +\infty}\frac{f(x)}{x}.
$$
Let us introduce the notation $\omega_f:=\lim_{x\to +\infty}\frac{f(x)}{x}.$
\par
So, according to \cite[Def. 2.1]{HMPB}, $f$ is regular enough to define the quantum $f$-divergence of positive semidefinite operators acting on a finite dimensional Hilbert space. The definition is the following.

\begin{definition}[\cite{HMPB}, Def. 2.1]
Let $f: [0,\infty) \rightarrow \R$ be a real-valued function on $[0, \infty)$ such that $f$ is continuous on $(0, \infty)$ and the limit $\omega_f=\lim_{x\to +\infty}\frac{f(x)}{x}$ exists in $[-\infty, +\infty].$ 
Let $A$ and $B$ be positive semidefinite operators on a finite dimensional Hilbert space $\cH.$
The \emph{quantum $f$-divergence} of A with respect to $B$ is defined as
\be \label{def0}
S_f(A||B):=\inner{B^\fel}{f\ler{L_A R_B^{-1}}B^\fel}_{HS}
\ee
if $\supp{A} \subseteq \supp{B}.$ If $\supp{A} \nsubseteq \supp{B},$ we define

\be \label{def1}
S_f(A||B):=\lim_{\eps\to 0} S_f(A||B+\eps I).
\ee

\end{definition}

\subsubsection{Computation rules}
A rather complicated computation shows that the quantum $f$-divergence of the positive semidefinite operators $A=\sum_{a \in \sigma(A)} a P_a$ and $B=\sum_{b \in \sigma(B)} b Q_b$ can be computed by the formula
\be \label{comp2}
S_f\ler{A || B}=\sum_{a \in \sigma(A)}\ler{\omega_f a \tr P_a Q_0+ \sum_{b \in \sigma(B)\setminus\{0\}} b f \ler{\frac
{a}{b}}\tr P_a Q_b},
\ee
see \cite[Cor. 2.3]{HMPB}.
It can be immediately seen from \eqref{comp2} that
\be \label{comp1}
S_f\ler{\lambda A|| A}= f(\lambda)\tr A \qquad \ler{A \in \cB(\cH)^+, \, \lambda \in [0, \infty)}.
\ee
\subsection{Examples}
\begin{example}
For the standard entropy function
$$
f(x)=
\begin{cases} 
x \log{x}, & \text{ if } x>0, \\
0, & \text{ if } x=0,
\end{cases}
$$
the induced quantum $f$-divergence on positive semidefinite matrices is the \emph{Umegaki relative entopy}
$$
S_f(A||B)=
\begin{cases}
\tr A \ler{\log{A}-\log{B}}, & \text{ if } \supp{A} \subseteq \supp{B}, \\
+\infty, & \text{ if } \supp{A} \nsubseteq \supp{B},
\end{cases}
$$
which is one of the most important numerical quantities in quantum information theory. Therefore, quantum $f$-divergences may be considered as generalized relative entropies \cite{HMPB}.
\end{example}
\begin{example}
For any $q>0, q\neq 1$ the function $f_q: x\mapsto f_q(x):=\frac{x^q-x}{q-1}$ is strictly convex, and the induced quantum $f$-divergence is
$$
S_{f_q}\ler{A||B}=
\begin{cases}
\frac{1}{q-1}\tr \ler{A^q B^{1-q}-A}, & \text{ if } \supp{A} \subseteq \supp{B} \text{ or } q<1,\\
+\infty, & \text{ if } \supp{A} \nsubseteq \supp{B} \text{ and } q>1,
\end{cases}
$$
which is the \emph{Tsallis relative entropy} of $A$ and $B$ if $0<q<1$ --- see, e. g., \cite{abe} or \cite{furu}.
\end{example}

\begin{example}
For the strictly convex function $f(x)=\ler{\sqrt{x}-1}^2$ the induced quantum $f$-divergence is
$$
S_f(A||B)=\norm{\sqrt{A}-\sqrt{B}}_{HS}^2
$$
see \cite{mnsz13}.
\end{example}

\section{The main results}
It is clear that any unitary or antiunitary conjugation leaves the quantum $f$-divergences invariant.
The main result of this paper is that the converse statement is also true, i. e., for any strictly convex function $f,$ the preservers of quantum $f$-divergences are necessarily unitary or antiunitary conjugations.

\begin{Remark}
Let $f$ be an arbitrary affine function, that is, $f(x)=\alpha x + \beta$ for some $\alpha, \beta \in \R.$
Easy computations show that in this case we have
$$
S_f \ler{A || B}=\alpha \tr A +\beta \tr B \qquad \ler{A, B \in \cB(\cH)^+}.
$$
So the quantum $f$-divergence of $A$ and $B$ depends only on the traces of the operators. Consequently, any trace preserving transformation of $\cB(\cH)^+$ preserves the quantum $f$-divergence, as well. This case is clearly out of our interest. Therefore, we investigate only those quantum $f$-divergences which are generated by strictly convex functions.
\end{Remark}

The precise formulation of our main statement is as follows.

\begin{theorem} \label{fo}
Let $f: [0, \infty) \rightarrow \R$ be a strictly convex function.
Let $\phi: \cB(\cH)^+ \rightarrow \cB(\cH)^+$ be a bijection which preserves the quantum $f$-divergence, that is,
$$
S_f(\phi(A)||\phi(B))=S_f(A||B) \qquad \ler{A, B \in \cB(\cH)^+}.
$$
Then there exists a unitary or antiunitary transformation $U: \cH \rightarrow \cH$ such that
$$
\phi(A)=UAU^* \qquad \ler{A \in \cB(\cH)^+}.
$$
\end{theorem}

\section{Proofs}

\subsection{The proof of Theorem \ref{fo}}

Let $f: [0, \infty) \rightarrow \R$ be a strictly convex function and assume that $\phi: \cB(\cH)^+ \rightarrow \cB(\cH)^+$ is a bijection which preserves the quantum $f$-divergence, that is,
$$
S_f(\phi(A)||\phi(B))=S_f(A||B) \qquad \ler{A, B \in \cB(\cH)^+}.
$$
Our aim is to prove that there exists a unitary or antiunitary transformation $U: \cH \rightarrow \cH$ such that
$$
\phi(A)=UAU^* \qquad \ler{A \in \cB(\cH)^+}.
$$
The proof is divided into two steps.

\begin{step}
Any quantum $f$-divergence preserving bijective transformation $\phi: \cB(\cH)^+ \rightarrow \cB(\cH)^+$ preserves the trace. 
\end{step}
\begin{proof}
By the computation rule \eqref{comp1},
$$
S_f\ler{A||A}=f(1)\tr A \qquad \ler{A \in \cB(\cH)^+}.
$$
If $\phi$ preserves the $f$-divergence, then
\be \label{egyes}
f(1)\tr \phi(A)=S_f\ler{\phi(A)||\phi(A)}=S_f\ler{A||A}=f(1) \tr A \qquad \ler{A \in \cB(\cH)^+}.
\ee
So if $f(1)\neq 0$ then the equation \eqref{egyes} immediately implies that $\phi$ preserves the trace, that is,
$$
\tr \phi(A)=\tr A \qquad \ler{A \in \cB(\cH)^+}.
$$
From now, throughout this proof we assume that $f(1)=0.$ Let us consider the following two possibilities.
\par
\emph{First case:} $f$ is strictly monotone decreasing. In this case $\omega_f=\lim_{x\to +\infty}\frac{f(x)}{x}\leq 0$ as $f(x)<0$ for any $x>1.$
Assume that the positive semidefinite operators $X$ and $A$ admit the spectral decompositions $X=\sum_{x \in \sigma(X)} x P_x$ and $A=\sum_{a \in \sigma(A)} a Q_a.$ By \eqref{comp2},
$$
S_f\ler{X||A}=\sum_{x \in \sigma(X)}\ler{\omega_f x \tr P_x Q_0+ \sum_{a \in \sigma(A)\setminus\{0\}} a f \ler{\frac
{x}{a}}\tr P_x Q_a}
$$
$$
\leq
\sum_{x \in \sigma(X)}\ler{0 \cdot x \tr P_x Q_0+ \sum_{a \in \sigma(A)\setminus\{0\}} a f \ler{0}\tr P_x Q_a}=f(0)\tr A.
$$
We used that in the above expressions $x \tr P_x Q_0 \geq 0$ and $a \tr P_x Q_a \geq 0$ as the trace of the product of two projections is nonnegative. On the other hand, by \eqref{comp1},
$$
S_f\ler{0||A}=f(0) \tr A.
$$
So 
$$
\max_{X \in \cB(\cH)^+} S_f(X||A)= f(0) \tr A.
$$
By the bijectivity and the $f$-divergence preserving property of $\phi,$ one gets
$$
f(0) \tr A= \max_{X \in \cB(\cH)^+} S_f(X||A)=\max_{X \in \cB(\cH)^+} S_f\ler{\phi(X)||\phi(A)}
$$
$$
=\max_{Y \in \cB(\cH)^+} S_f\ler{Y||\phi(A)}= f(0) \tr \phi(A).
$$
Clearly, $f(0)\neq 0$ as $f(0)>f(1)=0$ by the strict monotonicity of $f.$ Therefore, $\tr \phi(A)=\tr A.$

\par
\emph{Second case:} $f$ is not strictly monotone decreasing, that is, $f\ler{\gamma '}\leq f\ler{\delta}$ for some $0\leq \gamma ' <\delta.$ By the strict monotonicity of the difference quotient function $h(a,b)=\frac{f(a)-f(b)}{a-b}$ (Proposition \ref{convmon}), for any $\gamma \in \R$ which satisfies $0 \leq \gamma '<\gamma<\delta$ we have
$$
\frac{f(\delta)-f(\gamma)}{\delta-\gamma}>\frac{f(\delta)-f(\gamma ')}{\delta-\gamma '} \geq 0.
$$
So, $f(\gamma)<f(\delta)$ for some $0<\gamma<\delta.$ Let us make a few observations.
\bit
\item 
The function $f$ is continuous on the closed interval $[\gamma,\delta] \subset (0, \infty),$ hence it is bounded on $[\gamma,\delta].$
\item
By Proposition \ref{convmon}, the inequality
$$
\frac{f(\delta)-f(\beta)}{\delta-\beta}<\frac{f(\delta)-f(\gamma)}{\delta-\gamma}
$$
holds for any $\beta \in [0, \gamma).$ The above inequality is equivalent to
$$
f(\beta)>f(\delta)+\frac{\delta-\beta}{\delta-\gamma}\ler{f(\gamma)-f(\delta)}.
$$
Clearly, the right hand side is bounded from below as $\beta$ runs through the interval $[0, \gamma).$
Therefore, $f$ is bounded from below on $[0, \gamma).$
\item
Let us use again the statement of Proposition \ref{convmon}.
For any $\eps>\delta$ one has $f(\eps)>f(\gamma)$ as
$$
\frac{f(\eps)-f(\gamma)}{\eps-\gamma}>\frac{f(\delta)-f(\gamma)}{\delta-\gamma} > 0.
$$
So $f$ is bounded from below on $(\delta, \infty).$
\eit
By the above three observations, $f$ is bounded from below, hence $f$ has a finite infimum. This infimum is not necessarily a minimum. For example, the strictly convex function
$$
f(x)=\begin{cases}
      x^2-1, & \text{ if } x>0, \\
      0, & \text{ if } x=0
     \end{cases}
$$
does not have a minimum on $[0, \infty).$
\par
Let us introduce the notation $K:=\inf_{x \in [0, \infty)}f(x).$ We intend to show that
$$\inf_{X \in \cB(\cH)^+}S_f(X||A)=K \tr A.$$

The inequality $\inf_{X \in \cB(\cH)^+}S_f(X||A)\leq K \tr A$ is an immediate consequence of the computation rule \eqref{comp1} which states that $S_f\ler{\lambda A|| A}= f(\lambda)\tr A$ for any $A \in \cB(\cH)^+$ and $\lambda\geq 0.$
On the other hand --- using that $\omega_f \geq 0$ as $f$ is bounded from below --- for any $X=\sum_{x \in \sigma(X)}x P_x \in \cB(\cH)^+$ we have
$$
S_f\ler{X||A}=\sum_{x \in \sigma(X)}\ler{\omega_f x \tr P_x Q_0+ \sum_{a \in \sigma(A)\setminus\{0\}} a f \ler{\frac
{x}{a}}\tr P_x Q_a}
$$
$$
\geq
\sum_{x \in \sigma(X)}\ler{0 \cdot x \tr P_x Q_0+ \sum_{a \in \sigma(A)\setminus\{0\}} a K\tr P_x Q_a}=K\tr A,
$$
hence $\inf_{X \in \cB(\cH)^+}S_f(X||A)\geq K \tr A.$ (The operator $A$ is assumed to admit the spectral decomposition $A=\sum_{a \in \sigma(A)} a Q_a.$)
The transformation $\phi$ is bijective and preserves the quantum $f$-divergence, hence
$$
K \tr A= \inf_{X \in \cB(\cH)^+} S_f(X||A)=\inf_{X \in \cB(\cH)^+} S_f\ler{\phi(X)||\phi(A)}
$$
$$
=\inf_{Y \in \cB(\cH)^+} S_f\ler{Y||\phi(A)}= K \tr \phi(A).
$$
Note that $K\leq 0$ as $f(1)=0.$ If $K< 0,$ then the above equation clearly implies that $\tr \phi(A)=\tr A.$
\par
If $K=0,$ then $f(x)>0$ for any $x \neq 1,$ as $f$ is strictly convex. Moreover, the strict convexity of $f$ immediately implies that $f$ is strictly monotone decreasing on $[0,1].$ By Proposition \ref{convmon}, $\frac{f(x)-f(1)}{x-1}=\frac{f(x)}{x-1}$ is strictly monotone increasing (and positive) on $(1, \infty),$ and so is the function $x \mapsto \frac{x-1}{x}.$ The product of two strictly monotone increasing positive functions is strictly monotone increasing, hence $\frac{f(x)}{x}=\frac{f(x)}{x-1}\frac{x-1}{x}$ is strictly monotone increasing on $(1, \infty).$
In particular,
\be \label{moncsi}
\frac{b}{a} f \ler{\frac{a}{b}}<\omega_f \text{ for any } 0<b<a. 
\ee
We intend to prove the following characterization of the element $0 \in \cB(\cH)^+.$ Let $X \in \cB(\cH)^+.$ Then
$X=0$ if and only if 
$$
S_f\ler{A||B} \leq S_f\ler{A||X}+S_f \ler{X||B}
$$
for every $A, B \in \cB(\cH)^+.$
\par
Assume that $X  \neq 0.$ Consequently, $\tr X >0.$ By Proposition \ref{convmon},
$$
\frac{f(\mu)-f(1)}{\mu-1}<\frac{f(\lambda \mu)-f(\lambda)}{\lambda(\mu-1)}
$$
for any $\lambda>1$ and $\mu>1.$
Equivalently,
$$
\lambda f(\mu)+f(\lambda)< f(\lambda \mu).
$$
Multiplying with $\tr X$ we get that
$$
 f(\mu) \tr \lambda X +f(\lambda) \tr X< f\ler{\lambda \mu} \tr X
$$
which is equivalent by \eqref{comp1} to
$$
S_f\ler{\lambda\mu X||\lambda X}+S_f\ler{\lambda X||X} < S_f\ler{ \lambda\mu X||X}. 
$$
To see the converse statement, we need to show that $S_f\ler{A||B} \leq S_f\ler{A||0}+S_f \ler{0||B}$ for any $A, B \in \cB(\cH)^+.$ By the computation rule \eqref{comp2}, if $A=\sum_{a \in \sigma(A)} a P_a$ and $B=\sum_{b \in \sigma(B)} b Q_b,$ then
$$
S_f\ler{A || B}=\sum_{a \in \sigma(A)}\ler{\omega_f a \tr P_a Q_0+ \sum_{b \in \sigma(B)\setminus\{0\}} b f \ler{\frac
{a}{b}}\tr P_a Q_b}
$$
$$
=\sum_{a \in \sigma(A)} \omega_f a \tr P_a Q_0
+\sum_{\{(a,b)| a \in \sigma(A), b \in \sigma(B)\setminus\{0\}, a<b\}} b f \ler{\frac{a}{b}}\tr P_a Q_b
$$
$$
+\sum_{\{(a,b)| a \in \sigma(A), b \in \sigma(B)\setminus\{0\}, a=b\}} b f \ler{\frac{a}{b}}\tr P_a Q_b
$$
$$
+\sum_{\{(a,b)| a \in \sigma(A), b \in \sigma(B)\setminus\{0\}, a>b\}} b f \ler{\frac{a}{b}}\tr P_a Q_b.
$$
The term
$$
\sum_{\{(a,b)| a \in \sigma(A), b \in \sigma(B)\setminus\{0\}, a=b\}} b f \ler{\frac{a}{b}}\tr P_a Q_b
$$
is zero as $f(1)=0.$ Furthermore,
$$
\sum_{\{(a,b)| a \in \sigma(A), b \in \sigma(B)\setminus\{0\}, a<b\}} b f \ler{\frac{a}{b}}\tr P_a Q_b
$$
$$
\leq \sum_{\{(a,b)| a \in \sigma(A), b \in \sigma(B)\setminus\{0\}, a<b\}} b f(0) \tr P_a Q_b
$$
$$
=f(0)\sum_{b \in \sigma(B)} b \tr Q_b \ler{\sum_{a \in \sigma(A), a<b} P_a}\leq f(0)\sum_{b \in \sigma(B)} b \tr Q_b=f(0) \tr B,
$$
because $f$ is strictly monotone decreasing on $[0,1].$ By \eqref{moncsi},
$$
\sum_{a \in \sigma(A)} \omega_f a \tr P_a Q_0 +\sum_{\{(a,b)| a \in \sigma(A), b \in \sigma(B)\setminus\{0\}, a>b\}} b f \ler{\frac{a}{b}}\tr P_a Q_b
$$
$$
=
\sum_{a \in \sigma(A)} \omega_f a \tr P_a Q_0 +\sum_{\{(a,b)| a \in \sigma(A), b \in \sigma(B)\setminus\{0\}, a>b\}} a \frac{b}{a} f \ler{\frac{a}{b}}\tr P_a Q_b
$$
$$
\leq
\sum_{a \in \sigma(A)} \omega_f a \tr P_a Q_0 +\sum_{\{(a,b)| a \in \sigma(A), b \in \sigma(B)\setminus\{0\}, a>b\}} a \omega_f \tr P_a Q_b
$$
$$
=\sum_{\{(a,b)| a \in \sigma(A), b \in \sigma(B), a>b\}} a \omega_f \tr P_a Q_b=\omega_f \sum_{a \in \sigma(A)} a \tr P_a \ler{\sum_{b \in \sigma(b), b<a} Q_b}
$$
$$
\leq \omega_f \sum_{a \in \sigma(A)} a \tr P_a=\omega_f \tr A.
$$
We deduced that $S_f\ler{A || B} \leq f(0) \tr B+\omega_f \tr A.$ On the other hand, \eqref{comp2} shows that
$$
S_f(A||0)=\omega_f \tr A \text{ and } S_f(0||B)=f(0) \tr B.
$$
That is,
$$
S_f\ler{A || B} \leq S_f(A||0)+S_f(0||B).
$$
Now we can easily deduce that $\phi(0)=0.$ If $\phi$ preserves the $f$-divergence, then
$$
S_f\ler{\phi(A)||\phi(B)} \leq S_f\ler{\phi(A)||\phi(0)}+S_f \ler{\phi(0)||\phi(B)}
$$
holds for any $A, B \in \cB(\cH)^+.$ The map $\phi$ is bijective, hence this implies that $\phi(0)=0.$
The function $f$ is strictly monotone decreasing on $[0,1]$ and $f(1)=0,$ so $f(0)>0.$
By the computation rule \eqref{comp1} and by the preserver property of $\phi,$ the equation
$$
f(0) \tr A=S_f\ler{0||A}=S_f\ler{\phi(0)||\phi(A)}=S_f\ler{0||\phi(A)}=f(0) \tr \phi(A)
$$
holds for any $A \in \cB(\cH)^+.$ Consequently, $\tr \phi(A)=\tr A,$ that is, $\phi$ preserves the trace.
\end{proof}

\begin{step}
If a bijection $\phi: \cB(\cH)^+\rightarrow \cB(\cH)^+$ preserves the quantum $f$-divergence and the trace, then it is implemented by a unitary or an antiunitary operator.
\end{step}
\begin{proof}
Let us introduce the notation
$$
\cB(\cH)_\lambda^+ :=\left\{A \in \cB(\cH)^+: \, \tr A=\lambda\right\}.
$$
Note that $\cB(\cH)_1^+$ is nothing else but the \emph{state space} and it is denoted by $\cS(\cH).$ Observe that by the trace preserving property, $\phi$ restricted to $\cB(\cH)_\lambda^+$ is a bijection from $\cB(\cH)_\lambda^+$ onto itself.
It can be easily seen by the computation rule \eqref{comp2} that the quantum $f$-divergence is homogeneous, that is,
$$
S_f\ler{\lambda A || \lambda B}= \lambda S_f\ler{A || B} \qquad \ler{A, B \in \cB(\cH)^+, \, \lambda \in [0, \infty)}.
$$
For any $\lambda \in (0,\infty),$ let us define a map $\psi_\lambda$ in the following way:
$$
\psi_\lambda: \cS(\cH) \rightarrow \cS(\cH), \quad A \mapsto \psi_\lambda(A):=\frac{1}{\lambda}\phi\ler{\lambda A}.
$$
$\psi_\lambda$ preserves the $f$-divergence, because
$$
S_f\ler{\psi_\lambda(A)||\psi_\lambda(B)}
=S_f\ler{\frac{1}{\lambda}\phi\ler{\lambda A}||\frac{1}{\lambda}\phi\ler{\lambda B}}
$$
$$
=\frac{1}{\lambda} S_f\ler{\phi\ler{\lambda A}||\phi\ler{\lambda B}}
=\frac{1}{\lambda} S_f\ler{\lambda A||\lambda B}=S_f\ler{A||B}
$$
holds for any $A, B \in \cS(\cH)$ and $\lambda \in (0, \infty).$
Therefore, by the result of Moln\'ar, Nagy and Szokol \cite[Theorem]{mnsz13},
$$
\psi_\lambda(A)=U_\lambda A U_\lambda^*
$$
for some unitary or antiunitary operator $U_\lambda$ acting on $\cH.$ Let $P \in \cP_1(\cH)$ and $\lambda, \mu >0$ be arbitrary. Clearly,
$$
S_f\ler{\lambda P || \mu P}=\mu f\ler{\frac{\lambda}{\mu}}.
$$
On the other hand, for any positive $\lambda$ and $\mu,$
$$
S_f\ler{\phi\ler{\lambda P} || \phi\ler{\mu P}}
=S_f\ler{\lambda \psi_\lambda (P) ||\mu \psi_\mu (P)}
$$
$$
=\mu f\ler{\frac{\lambda}{\mu}} \tr \psi_\lambda (P) \psi_\mu (P)
$$
$$
+\lambda \omega_f \tr \ler{I-\psi_\mu (P)}\psi_\lambda (P)+\mu f(0) \tr \ler{I-\psi_\lambda (P)}\psi_\mu (P).
$$
The operators $\psi_\lambda (P)$ and $\psi_\mu (P)$ are rank-one projections, so their traces are equal to one. Using this fact and the $f$-divergence preserving property of $\phi,$ we get that
$$
0=S_f\ler{\phi\ler{\lambda P} || \phi\ler{\mu P}}-S_f\ler{\lambda P || \mu P}
$$
$$
=\ler{1-\tr \psi_\lambda (P) \psi_\mu (P)}\ler{\lambda \omega_f+\mu f(0)}-\ler{1-\tr \psi_\lambda (P) \psi_\mu (P)}\mu f\ler{\frac{\lambda}{\mu}}
$$
$$
=\ler{1-\tr \psi_\lambda (P) \psi_\mu (P)}\ler{\lambda \omega_f+\mu f(0)-\mu f\ler{\frac{\lambda}{\mu}}}.
$$
The term $\lambda \omega_f+\mu f(0)-\mu f\ler{\frac{\lambda}{\mu}}$ is stricty positive, as by the strict monotonicity of the difference quotient function (Proposition \ref{convmon})
$$
\frac{f\ler{\frac{\lambda}{\mu}}-f(0)}{\frac{\lambda}{\mu}-0}<\omega_f
$$
for any $0<\lambda, \mu.$ Therefore, $\tr \psi_\lambda (P) \psi_\mu (P)=1,$ hence $\psi_\lambda (P) =\psi_\mu (P).$
We know that both $\psi_\lambda$ and $\psi_\mu$ are affine maps and any element of $\cS(\cH)$ is a convex combination of rank-one projections, so $\psi_\lambda=\psi_\mu$ for any $\lambda$ and $\mu.$
This means that there is a unitary or antiunitary operator $U$ such that $\phi(A)=U A U^* \, \ler{A \in \cB(\cH)^+}.$
\end{proof}

We close this paper with the following
\begin{Remark}
In contrast with the result of \cite{mnsz13}, our Theorem concerns only bijective transformations of the cone of positive semidefinite operators.
The description of non-bijective transformations which leave quantum $f$-divergences invariant seems to be a challenging problem.
We propose it as an open question.
\end{Remark}

\subsection*{Acknowledgement}
The author is grateful to Lajos Moln\'ar for comments and suggestions which helped to improve this paper a lot.

\bibliographystyle{amsplain}

\end{document}